\documentclass{aa}
\usepackage{graphicx}
\usepackage{epsfig,times}
\usepackage{color}
\usepackage{amssymb}
\usepackage{natbib}
\bibpunct{(}{)}{;}{a}{}{,} 

\begin{document}

\def\eg{{\em e.g.}}
\def\ev{\,{\rm eV}}
\def\mev{\,{\rm MeV}}
\def\gev{\,{\rm GeV}}
\def\erg{\,{\rm erg}}
\def\ie{{\em i.e.}}
\def\km{\,{\rm km}}
\def\cm{\,{\rm cm}}
\def\wrt{{\it w.r.t.}}

\title{Quark-nova explosion inside a collapsar: \\ application
 to Gamma Ray Bursts}

\author{Rachid Ouyed, Denis Leahy,  Jan Staff, and Brian Niebergal}

\institute{Department of Physics and Astronomy, University of Calgary, 
2500 University Drive NW, Calgary, Alberta, T2N 1N4 Canada\thanks{email:ouyed@phas.ucalgary.ca}}

\date{Received <date>; accepted <date> }

\authorrunning{Ouyed et al.}

\titlerunning{Quark-Nova inside a collapsar}

\abstract{If a quark-nova occurs
 inside a collapsar, the interaction
  between the quark-nova ejecta (relativistic iron-rich chunks)  and the collapsar envelope,
   leads to features indicative of those observed in Gamma Ray Bursts.  
     The quark-nova ejecta collides with the stellar envelope creating an outward moving cap
      ($\Gamma\sim$ 1-10) above the polar funnel.  Prompt gamma-ray burst emission
       from internal shocks in relativistic jets (following accretion
        onto the quark star) become visible after  the cap  becomes optically thin.
Model features include: (i) precursor  activity (optical, X-ray, $\gamma$-ray), 
 (ii) prompt  $\gamma$-ray emission, and (iii) afterglow emission.
  We discuss SN-less long duration GRBs, short hard GRBs (including association and
  non-association with star forming regions),  
   dark GRBs,  the energetic  X-ray
 flares detected in  Swift GRBs, and the 
  near-simultaneous optical and $\gamma$-ray prompt emission observed in GRBs
   in the context of our model.
}

\maketitle

\section{INTRODUCTION}

Recent observations following the launch of the {\it Swift} satellite challenge the traditional
 models of GRBs (e.g. M\'esz\'aros  2006; Zhang 2008). 
In particular the traditional afterglow modeling, 
 which has been successful in many ways,  appears to have serious limitations (e.g. Granot 2008 for a recent
 review).
 Here we  show how appealing to a quark-nova occurring inside a collapsar can lead
 to phenomonology reminiscent of that seen by Swift. We start with a brief review of the Quark-Nova
explosion.

 A quark-nova (QN), (Ouyed, Dey, \& Dey 2002; Ker\"anen\&Ouyed 2003;
 Ker\"anen, Ouyed, \& Jaikumar 2005)
is the explosion driven by phase transition of the 
core of a neutron star (NS) to the quark matter phase (i.e neutron star core collapse) leading
 to the formation of a quark star (QS). 
The gravitational potential energy
released (plus latent heat of
phase transition) during this event is converted partly into internal energy  and partly into outward propagating shock waves
which impart kinetic energy to the material that forms the
ejecta (i.e. the outermost layers of the neutron star crust).
The ejection of the outer layers
 of the NS is driven by  
 the thermal fireball generated as the star cools from its birth temperature
 down to $\sim 7.7$ MeV (Vogt, Rapp, \& Ouyed 2004; Ouyed, Rapp, \& Vogt 2005).
The fireball
expands approximately adiabatically while pushing the overlaying crust, and cooling fairly rapidly. The energy needed to eject
the crust is less  than 1\% of fireball energy.

 The initial composition of the ejecta  is representative of matter in the outer layers of the neutron star crust
 (with density below $\sim 10^{11}$ g cm$^{-3}$), dominated by iron-group elements and neutron-rich
 large Z nuclei beyond iron (Baym, Pethick, \& Sutherland 1971). 
 As the ejecta expands, r-process takes effect leading to the formation of even heavier
elements.  As shown in Jaikumar et al. (2007), the QN is effective at turning at most 10\% of the ejecta
 into  elements  above  $A\sim 130$ (Jaikumar et al. 2007).
  In previous work we explored the dynamical and thermal evolution of this 
  ejecta  (Ouyed\&Leahy 2008; Leahy\&Ouyed 2008a). 
  As the ejecta moves outwards it expands and cools undergoing a liquid
   to solid transformation\footnote{For ejecta birth temperature of the order of 10 MeV, the relativistic electrons  are only mildly degenerate
 (see appendix in Ouyed\&Leahy 2008). 
 Thus additional heat deposition into the ejecta during the expansion (e.g. due to nuclear decays
 of r-processed elements) could lead to non-degeneracy leaving the ejecta in gaseous form.
 Here, we assume that the degeneracy is not lifted during the early ejecta expansion.}. The relativistic expansion causes rapid breakup into small chunks 
  because of the inability of causal communication laterally in the shell.
Whether liquid or solid iron, inter-ionic forces (mediated by the electrons) provide the tension leading to breakup (which does not occur for a gas).
    The size of the clumps depends
     on whether the breakup occurs in the liquid or solid phase.  In the solid/liquid
     phase the ejecta breaks up into $\sim 10^7$/$10^3$ chunks with chunk mass 
       of $\sim 10^{19}$/$10^{23}$ gm.
     Table 1 in Ouyed\&Leahy (2008) lists the properties of the clumps/chunks.

\begin{figure*}[t!]
\centering
\centerline{\includegraphics[width=1.0\textwidth,angle=0]{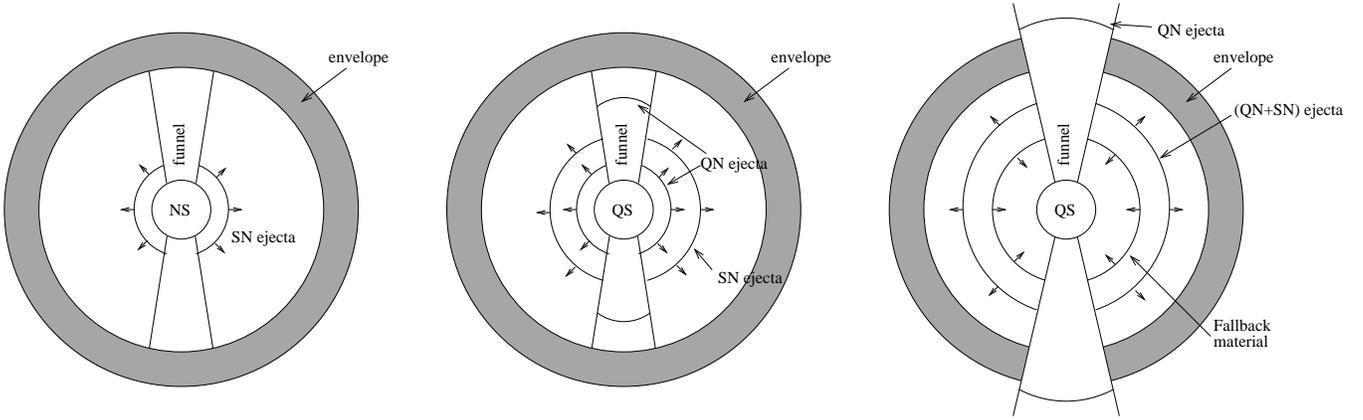}} 
\caption{\label{fig:intro}
Outline of the initial phases for GRBs from Quark-Novae (QN). {\it Stage 1}:
 A newly formed neutron star with expanding ejecta and SN shock wave,
as well as the stationary WR stellar envelope is also shown. Angular momentum of the progenitor
  results in low density polar funnels. {\it Stage 2}: An explosive neutron star to quark star conversion
   (i.e. Quark-Nova) occurs producing the QN ejecta. The QN ejecta can then propagate freely through the funnel,
    while in other directions it will overtake the SN ejecta. {\it Stage 3}:
    The QN ejecta along the funnel interacts with the WR stellar envelope, 
    while the collision
     of the SN and QN ejecta lead to  an energized outgoing ejecta (suggestive
      of a hypernova).}
\end{figure*}

In this paper, we explore consequences of a quark nova  occuring during the supernova explosion in a rotating massive star.  Before the QN has occurred, one has the progenitor collapse -- much like
 in the collapsar picture except that in our case a QS is formed
 instead of a black hole (BH) (see Fig.~\ref{fig:intro}). 
  We note that for low angular momentum progenitors, the combination
 of a high NS core density at birth and, most likely, fall-back material
 would drive the proto-neutron star to a  black hole.
 High angular momentum progenitors (collapsars), will delay 
 the formation  of a black hole for three main reasons:
 (a) the progenitor's core 
 tends to shed more mass and angular momentum as it shrinks
 reducing central core mass and fall-back; 
 (b) high spin keeps the core density of the resulting neutron star from crossing the black
  hole formation limit; 
 (c)  high angular momentum in the material around the core reduces the
accretion rate onto the central object. The subsequent accretion
 onto the quark star  explains the prompt emission
 in our model (Ouyed et al. 2005).
The conversion from NS to QS depends on the NS central density
 at birth. As shown by Staff et al. (2006),
 spin-down leads to increase of core density and subsequent conversion. 
 Thus progenitor's angular momentum  does not mean
 the conversion to QS is unlikely, it  only affects the delay between the SN and QN.
 In summary, collapsars seem to provide favorable
 conditions for the QN to occur inside them. Furthermore,  
 the high angular momentum in the envelope leads to funnel formation which allows the QS jet to escape
 the envelope and the GRB to be visible.

  The paper is structured as follows: In section 2 we investigate the ejecta interaction
  with the stellar envelope for the two cases of thin and thick envelope. In section 3
   we apply our model to GRBs and explain how the interaction of the chunks
    with the stellar envelope can lead to precursor, prompt and afterglow emissions
    reminiscent of those observed in GRBs. A discussion is given in section 4 before
    concluding in section 5.

\section{Ejecta's Interaction with stellar envelope}\label{sec:stellar_env}

Wolf-Rayet stars can have extended envelopes, the profile of which depends on evolution
and metallicity. The evolutionary effects generally result in WN (nitrogen burning)
stars evolving into WC (carbon burning) stars with much smaller masses and radii due
to mass-loss (e.g. Meynet \& Maeder 2003; Heger et al. 2003). 
What is of interest here is the structure of the envelope at the time
of stellar collapse, which is not yet fully understood.
For simplicity,  we take the stellar structure of
a Helium Wolf-Rayet star (Petrovic et al. 2006) to be representative of the progenitor
and consider the low and high-metallicity cases (see their Figure 2). The main difference is that the
high metallicity star has an extended envelope with density $\sim (10^{-10}$-$10^{-9}$) g cm$^{-3}$
and a density inversion near the surface ($\sim 3R_{\odot}$ for a $24M_{\odot}$
star). In the low metallicity case, the star envelope cuts off sharply at $\sim 1.5R_{\odot}$.
      
When the broken pieces of ejecta impact this stellar envelope they undergo
a shock and become heated to a temperature 
\begin{equation}
\label{eq:1}
T_{\rm c}\sim \xi_{\rm s}\Gamma_{\rm i} \frac{A}{1+Z}m_{\rm H}c^{2}\ ,
\end{equation}
 where $\Gamma_{\rm i}$ is the Lorentz factor of the ejecta and $m_{\rm H}$
 the proton mass.   
  Equation above shows that the chunk temperature is insensitive to the presence of 
  heavier elements since $A/(1+Z)$ does not vary much. 
 Hereafter, and for simplicity, we assume an iron-dominated ejecta (i.e. A=56 and Z=26).
Here, the shock efficiency was roughly 
estimated to be $\xi_{\rm s}\sim (\rho_{\rm env.}/\rho_{\rm Fe})^{2}$,
where $\rho_{\rm env}$ is the envelope density at the shock radius. 
Noting that non-degenerate iron will vaporize if heated to $\ge 0.3$ eV (see CRC tables 2005
 for vaporization temperature of iron at normal density),
we then define a critical envelope density $\rho_{\rm env,c}\sim 10^{-5}$
g cm$^{-3}$  above which the chunks will be vaporized and 
lose considerable momentum.

One might argue that normal core collapse supernovae accompanied by quark novae, should
 be more energetic than the canonical $\sim 10^{51}$ ergs observed.
 However,  this depends on the delay between the SN and the QN.
If the delay between the QN and SN is long enough, the chunks  will
 not re-energize the SN. 
  The mean density in the envelope ($\bar{\rho}_{\rm env.}\propto M_{\rm env.}/R_{\rm env.}^3$
  in the simplest of cases) depends 
on (i) the progenitor's pre-collapse profile (which depends on evolution and metallicity) and (ii) 
 on the delay between the QN and SN. The longer the delay, 
 the smaller the envelope density when the chunks hit it. 
For a typical $5M_{\odot}$ envelope we find that $\bar{\rho}_{\rm env.}\sim \rho_{\rm env.,c}$
 is reached when the envelope is at $R_{\rm env.}\sim  10^{13}$ cm. In other words,
 for cases where the delay between the SN and QN exceeds a few days
(for SN ejecta velocity $\sim  1000$ km s$^{-1}$), the QN ejecta will encounter a thin envelope yielding
 weak interaction.
For more massive envelopes,  delays of the order of weeks are required for the 
  density to drop below critical; shorter delays  lead to 
     complete dissipation of the chunks  energizing the preceding
     supernova remnant. As shown in Leahy\&Ouyed (2008b),    
     this  can account for superluminous supernovae  such as SN 2006gy. 
 The density of the stellar envelope along the rotation axis 
 is also affected by rotation.For a rotating progenitor
the collapse proceeds fastest along the polar axis leaving a low density path 
called the funnel (Woosley\&Bloom 2006 and references therein), with opening half angle $\theta_{\rm f}$. 
If the QN ejecta propagated into this funnel, then it would encounter
negligible resistance (i.e. thin envelope case), while in other directions the QN ejecta would interact
with the higher density SN ejecta (see Fig.~\ref{fig:intro}). 
In these equatorial regions most of the QN energy is lost to energizing the SN ejecta and only
a fraction ($\sim 2.5\times 10^{49}\ {\rm erg}\ \eta_{0.1}\theta_{\rm f,0.1}^{2}E_{\rm QN,53}$) 
is directed into the funnel.
The bulk of the ejecta energy not entering the funnel, $\sim 10^{52}\ {\rm erg}\ \eta_{0.1}E_{\rm QN,53}$,
goes into re-energizing the SN ejecta and can result in a hypernova (see discussion
in \S~\ref{sec:hypernova}).

 \subsection{Thick envelope} \label{sec:thick_env}
           
If the envelope density is higher 
 than $\rho_{\rm env,c}$, 
the chunks will be vaporized upon impact leading to runaway dissipation
 and total merging of the ejecta with the envelope.  
 A significant
fraction, $m_{\rm env.}/(m_{\rm env.}+ m_{\rm ejecta})$, of the kinetic energy of the QN ejecta goes into
heating the envelope. In this case, the thermal energy of the combined ejecta and envelope (hereafter referred to as  the cap) is 
\begin{equation}\label{eq:etherm}
 E_{\rm th.}= \frac{m_{\rm env}}{m_{\rm env}+m_{\rm ejecta}} 
 \frac{\pi\theta_{\rm f}^2}{4\pi}\Gamma_{\rm i} m_{\rm ejecta}c^2\ ,
\end{equation}
where the resulting thermal energy per nucleus is, 
\begin{equation}\label{eq:tenv}
  kT_{\rm nuc}\sim \frac{m_{\rm ejecta}m_{\rm env.}}
     {(m_{\rm ejecta}+m_{\rm env})^2}\Gamma_{\rm i} \mu m_{\rm H}c^2 \ ,
\end{equation}
with $\mu$ the mean mass per nucleus. 
The maximum nucleus temperature, $T_{\rm nuc.,max}\sim 2.4\ {\rm GeV}\ \Gamma_{i, 10}\mu$, 
occurs when the envelope and ejecta masses are equal; $\Gamma_{i, 10}$ is the ejecta's
 initial Lorentz factor in units of 10. 
Note that, $\mu\sim 1$ whenever $T_{\rm env.}$ exceeds 1 MeV due to nuclear dissociation. 
 Thermalization with  $(e^+e^-)$  pair creation places an upper limit on
the electron temperature of $\sim 1$ MeV. Subsequent energy transfer from nuclei
to electrons contributes to further pair creation as the nuclei cool to 1 MeV.
If $kT_{\rm nuc}> 1$ MeV, then most of the ejecta's kinetic energy ends up as
$(e^+e^-)$ pairs. The end result would then be a cap rich in pairs.

\begin{figure}[t!]
\centerline{\includegraphics[width=0.5\textwidth, angle=0]{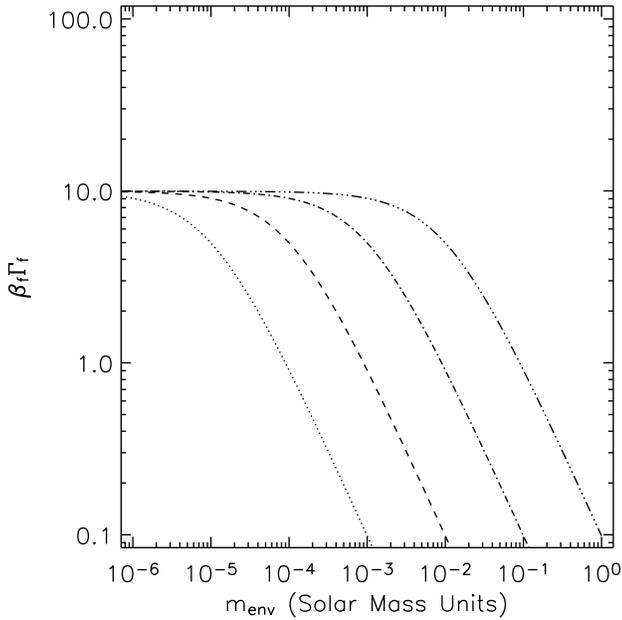}} 
\caption{\label{fig:gamma}
 Final velocity ($\beta_{\rm f}\Gamma_{\rm f}$; Eq.~\ref{eq:betagamma})
  of the combined mass of the envelope plus ejecta following collision.  The four curves
 are for ejecta masses ranging from $10^{-5}M_{\odot}$ to $10^{-2}M_{\odot}$, 
 left to right (all with a QN ejecta with $\Gamma_{\rm i}=10$). The  transition from relativistic
 to non-relativistic bulk motion occurs at $\beta_{\rm f}\Gamma_{\rm f}=1$.}
\end{figure}

Momentum conservation arguments in this case show that
the cap slows quickly, reaching a final speed ($v_{\rm f}$) of,
\begin{equation}\label{eq:betagamma}
  \beta_{\rm f}\Gamma_{\rm f} = \frac{\sqrt{\Gamma_{\rm i}^{2}-1}}{1+ \frac{m_{\rm env}}{m_{\rm ejecta}}}\ ,
\end{equation}  
where $\beta_{\rm f}=v_{\rm f}/c$. 
 We note that if the envelope mass is less than 
$m_{\rm env, R} \sim 10^{-3}M_{\odot} \Gamma_{\rm i, 10}m_{\rm ejecta,-4}$, then the mixed ejecta 
 is moving radially outwards at relativistic speeds 
($\beta_{\rm f}\Gamma_{\rm f}>1$ or $\beta_{\rm f}> 1/\sqrt{2}$; see Fig.~\ref{fig:gamma}).  
Thus $m_{\rm env, R}$ separates two regimes within the thick envelope case 
   which  is of relevance  when applying our model to GRBs.

    \subsection{Thin envelope}\label{sec:thin_env}

If the stellar envelope density following
the collapse is below the critical density, $\rho_{\rm env.,c}$, the chunks
will not be vaporized nor do they expand significantly, 
rather they pass through the envelope effectively puncturing it. 
During this interaction the temperature of a piece of broken ejecta, $T_{\rm c}$, 
 is determined by shock heating (eq.\ref{eq:1}), and 
 will not exceed the eV range;  thus any emission would be in the optical band (see \S \ref{sec:optical}).
 As discussed above, high metallicity stars can have extended thin envelopes.
 However, inhomogeneities and asphericity in the thick envelope case 
  could lead to low-density regions where the chunks can  survive destruction.

 \section{Application to GRBs}\label{sec:apply}

 The advent of the {\it Swift} mission has enabled a much more
intensive sampling of GRB light curves, particularly during
its early phases but also extending out to late times.
These data allow for a more stringent comparison with the
standard blast wave model. In addition to the suggested
 extended engine activity, the observed X-ray flares  (e.g. Nousek et al. 2006) appear to be a distinct emission component, which suggests
 a sporadic late time activity of the central source. 
Another interesting  finding by {\it Swift} is that the early optical emission, which has been
 attributed in some cases before {\it Swift} to the reverse shock, is typically dimmer than expected. 
 The chromatic breaks in the afterglow lightcurves is  puzzling as it  suggests that
 the X-ray and optical emission may arise in separate physical components, which would
 then naturally account for their seemingly decoupled lightcurves.
There are other features  that seem difficult to explain within the framework of the standard engine
 and afterglow  (we refer the interested
 reader to Granot 2008 for more on this).

Here we show how appealing to a quark-nova
 following the SN can help alleviate at least some of the issues 
 mentioned above. The interaction of the QN ejecta with the stellar envelope yields  
  precursors and postcursors in the optical and X-ray range as shown next.

 \subsection{Precursors}\label{sec:precursors}

\subsubsection{Thin envelope (optical precursor)}
\label{sec:optical}

In our model, the mechanism for production of an optical flash is the heating
 of the chunks of the QN ejecta in the thin envelope case.
    In this case, emission is directly
related to shock efficiency with emitted energy,
\begin{eqnarray}
E_{\rm p, O} &\sim&  (\xi_{\rm s} \Gamma_{\rm i} \mu_{\rm e} m_{\rm ejecta} c^2)\times \Gamma_{\rm i}^2\\\nonumber
 & \sim& 10^{41}~ {\rm ergs}\ \xi_{\rm s, c}\Gamma_{\rm i,10}^3 m_{\rm ejecta,-4}\ ,
\end{eqnarray}
  where $\xi_{\rm s,c}= (\rho_{\rm env.,c}/\rho_{\rm Fe})^2$ and $\mu_e\sim 2$ is the mean weight per electron.   
 The precursor optical emission we expect to be thermal-like despite such a relativistic beaming;
 equation above takes into account beaming correction ($\Gamma_{\rm i}^2\sim 100$).

 The precursor time is governed by 3 timescales:
 \begin{itemize}
\item (a) time to traverse the envelope, if the envelope is optically thin to the
radiation emitted at temperature $T_{\rm c}$; 
\item (b) geometrical time delay ($\theta_{\rm f} R_{\rm env}/c$);
\item (c) cooling time of the chunk:  $t_{\rm cool}\sim (3/2)kT_{\rm c} N_{\rm Fe}/L$
with $L=A_{\rm c} \sigma T_{\rm c}^4$;  the chunk's area is $A_{\rm c} \sim ~6\times 10^{11}$ cm$^2$.
\end{itemize}

The number of particles in the chunk is $N_{\rm Fe} \sim m_{\rm c}/(\mu m_{\rm p})\sim 2\times 10^{44}$ with
 $\mu\sim 28$ (if we take one ion and one free electron in the metal per Fe
nucleus). Then $t_{\rm cool}\sim 0.8\ {\rm s}\times
 T_{\rm keV}^{-3}$ 
and depends strongly on the temperature that the chunks are heated to.
If optical (eV) then {\it (c)} is longer than {\it (a)} so the longer of {\it (b)} and {\it (c)}
would give the observed precursor duration. If the chunks are heated to keV
temperatures, {\it (c)} is so short  that unless 
the chunk is continuously heated by interaction with the envelope, 
 the longer of {\it (a)} or {\it (b)} would give observed precursor duration.
The observed cooling time is shorter by a factor of $1/(2 \Gamma_{\rm c}^2$) due to
relativistic motion of the chunk toward the observer.
 We note that for liquid clumps $N_{\rm Fe}$ and the area $A_{\rm c}$ are both larger somewhat
lengthening $t_{\rm cool}$. But $t_{\rm cool}$ is still dominated by the
value of $T_{\rm keV}$ leading to the same conclusions.

Near-simultaneous optical and $\gamma$-ray  emission  has been observed
 in a few cases (e.g. Zou, Piran, \& Sari 2008). This has led to open debates on
  the association or non-association between the two emissions (e.g. Kumar\&Panaitescu 2008).
   This is further discuss in \S \ref{sec:opticalemission}. Let us simply mention that in our model, 
 any observation of optical precursor, means that the envelope
density must be  close to $\rho_{\rm env., c}$  with chunks heated
to 0.3 eV (i.e.  $T_{\rm observed}= \Gamma_{\rm i}\times 0.3\ {\rm eV}\sim 3$ eV). In those
sources, the observed optical precursor  could yield crucial
 information about  the delay between the SN and QN.

 \subsubsection{Thin envelope with density inversion (optical and X-ray precursors)}
 \label{sec:xray}
       
After the chunks  have freely propagated outside of the main envelope
they can interact with a higher density shell further out at a radius of $r_{\rm inv}$ (i.e. the density
inversion in the envelope at a few times $10^{11}$ cm; e.g. figure 2 in Petrovic et al. 2006). 
In order for the chunks to dissipate their energy, the density of the outer shell
must exceed $\rho_{\rm env,c}\sim 10^{-5}$ g cm$^{-3}$ (from \S~\ref{sec:stellar_env}).
Once the chunks  collide with matter possessing this critical density they spread,
resulting in their density decreasing, initiating a runaway dissipation process, ionization
 and heating.  The emission from the shocked material is optically thin 
 so the observer sees radiation  at the shock temperature $T_{\rm c}$
  as given in equation (\ref{eq:1}). For example, if the density at the inversion
   is $\sim 100\ \rho_{\rm env., c}$, then the X-ray emission will peak at $\sim 3$ keV.
   
   Since the mass at of the envelope at the inversion radius
    is much less than the ejecta mass, the shock propagates at
     $\sim \Gamma_{\rm i}$. Since $1/\Gamma_{\rm i} \ge \theta_{\rm f}$, 
    an observer would see the emission from all of the chunks, and so the 
overall precursor pulse would be due to the sequential viewing of different
individual pulses from each chunk along the curved surface.
The precursor duration  is then due to a geometrical delay,
\begin{equation}\label{eq:dur}
  t_{\rm prec.}\sim \frac{\theta_{\rm f}r_{\rm inv}}{c}\sim 3\ {\rm s}\  \theta_{\rm f,0.1}r_{\rm inv,12}\ .
\end{equation}

 \subsubsection{Thick envelope case ($\gamma$-ray Precursor)}
\label{sec:gamma}
         
If significant portions of the stellar envelope are above the critical density, 
then one would expect the kinetic energy 
of the chunks to be deposited in a thin dissipation zone at the base of the envelope.
This will effectively spread the ejecta, forming a piston
with a strong shock ahead of it.  This piston should remain
relativistic until it has swept up approximately $\Gamma_{\rm i} m_{\rm ejecta}$
of envelope mass, at which point it slows, reaching a final velocity given by
equation (\ref{eq:betagamma}).  Although the shock heats up and dissociate the nuclei, however, as noted above, the actual temperature\footnote{The details of the shock heating of the envelope and its subsequent
cooling are complex. Using blackbody cooling as an upper
limit leads to an extremely rapid cooling time (mainly due to the large emitting area) 
of $t_{\rm env,cool}\sim 10^{-12}\ {\rm s}\ m_{\rm env,-4}/T_{\rm env,1}^{3}$, 
where the envelope temperature is in units of 1 MeV.
          We note that the actual cooling time is defined by the shock
          propagation time through the optically thin outer parts of the envelope ($\sim 10^{5}$ cm), which
           yields timescales $\le 10^{-4}$ s. }
will be limited to $\sim 1$ MeV due to thermal $(e^+e^-)$ pair creation.

\begin{figure}[t!]
\centerline{\includegraphics[width=0.5\textwidth,angle=0]{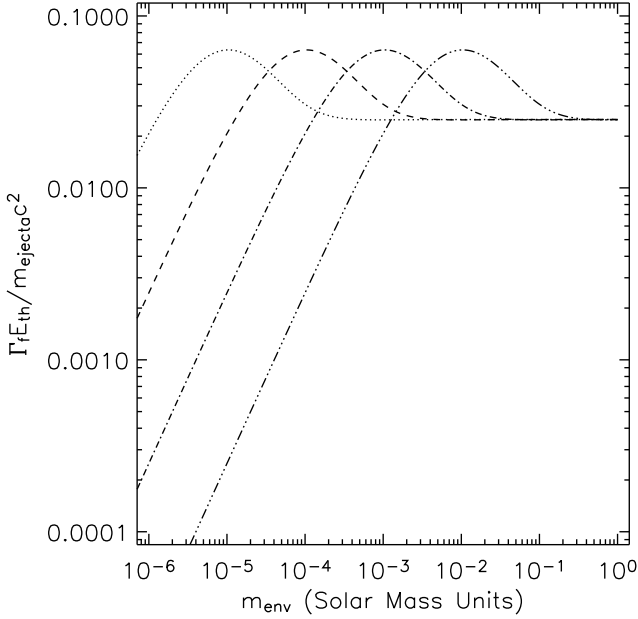}} 
\caption{\label{fig:fluence}
 Estimated fluence of the precursor in the thick envelope case (from Eq.~\ref{eq:etherm}). 
 The four curves
 are for ejecta masses from $10^{-5}M_{\odot}$ to $10^{-2}M_{\odot}$ from
 left to right (all with a QN ejecta with $\Gamma_{\rm i}=10$).  The peak fluence occurs at $m_{\rm ejecta}=2m_{\rm env}$
  with a value of about 6\% of $m_{\rm ejecta}c^2$.}
\end{figure} 

The precursor consists of a   short burst
of radiation when the shock reaches the outer edge of the envelope.
 The precursor  would have a typical temperature of a pair
plasma, $T_{\rm prec.}\sim  500$ keV with a 
 duration   also defined by geometrical delays,
\begin{equation}
 t_{\rm prec.} \sim \frac{\theta_{\rm f}r_{\rm env.}}{c} 
\sim  0.3\  {\rm s}\ \theta_{\rm f,0.1}r_{\rm env.,11} \ .
\end{equation}
The precursor brightness in the thick envelope case depends on the final speed
            of the combined ejecta: 
            (i) if it is relativistic ($m_{\rm env.} < m_{\rm env., R}$)  the usual blueshift and beaming
            applies yielding higher brightness ($\propto \Gamma^{2}$); (ii)
             if it is not relativistic we expect the precursor to be  dimmer and harder
             to detect.  
             
              We  approximate the precursor fluence by assuming that all of the
           thermal energy is radiated. In the non-relativistic case
           and for $m_{\rm env, c.} < m_{\rm env,R} < m_{\rm env}$ 
\begin{equation}
E_{\rm p} \sim  2.5\times 10^{49}\ {\rm ergs}\ \eta_{0.1} \theta_{\rm f,0.1}^{2}E_{\rm QN,53}\ ,
\end{equation} 
        while    the resulting fluence in the relativistic case (i.e. $m_{\rm env, c.} < m_{\rm env} < m_{\rm env,R}$) is written as  
          $\Gamma_{\rm f}E_{\rm th.}$ and  is shown in Figure (\ref{fig:fluence}). It 
            peaks at $ (2\Gamma_{\rm i}^{2}/9) (\theta_{\rm f}^2/4)m_{\rm ejecta}c^{2}$, or,
           \begin{equation}
           E_{\rm p, max} \sim  2.5\times 10^{50}\ {\rm erg}\  \frac{\theta_{\rm f,0.1}^{2} \eta_{0.1}^{2} E_{\rm QN,53}^{2}}{m_{\rm ejecta,-4}}\ .
           \end{equation}
             
             Note that when $m_{\rm ejecta}/m_{\rm env.}< 10^{-5}$, the thermal
             energy per nucleus is in the keV range leading to an X-ray precursor instead
               of a $\gamma$-ray precursor.

\begin{figure*}[t]
\centering
\includegraphics[scale=0.48]{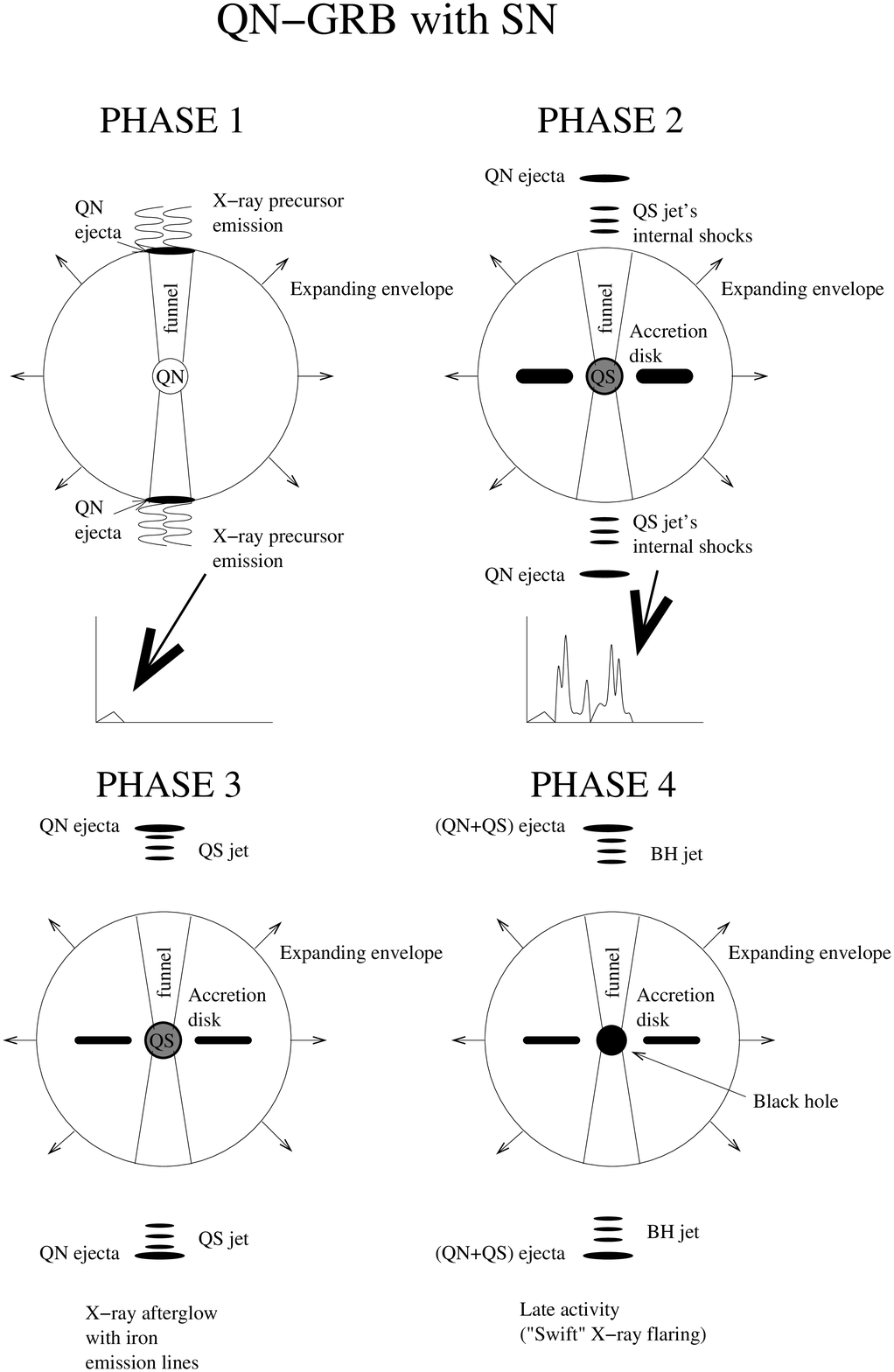}
\caption{\label{fig:sn}
Shown in this figure are three sequential phases and a fourth possible
 phase in our model for the case of a SN (i.e. expanding envelope).
  Phase 1  consists of the interaction between the QN ejecta and the envelope
   leading to an X-ray precursor (see \S \ref{sec:thin_env}). An accretion disk forms 
   around the QS leading to Phase 2 where a jet
    is launched, providing the prompt GRB emission (see \S \ref{sec:prompt_emiss}).  Phase 3 shows
     the late stages of the QS jet interacting with the QN ejecta leading to
     an X-ray afterglow.  However, if accretion is sufficiently large the QS
      may turn into a BH (phase 4 above), causing launch of a second jet extending the
       prompt GRB phase reminiscent of  late time activity observed by Swift (see
        Staff et al. 2006b)}
\end{figure*}

\begin{figure*}[t]
\centering
\includegraphics[scale=0.48]{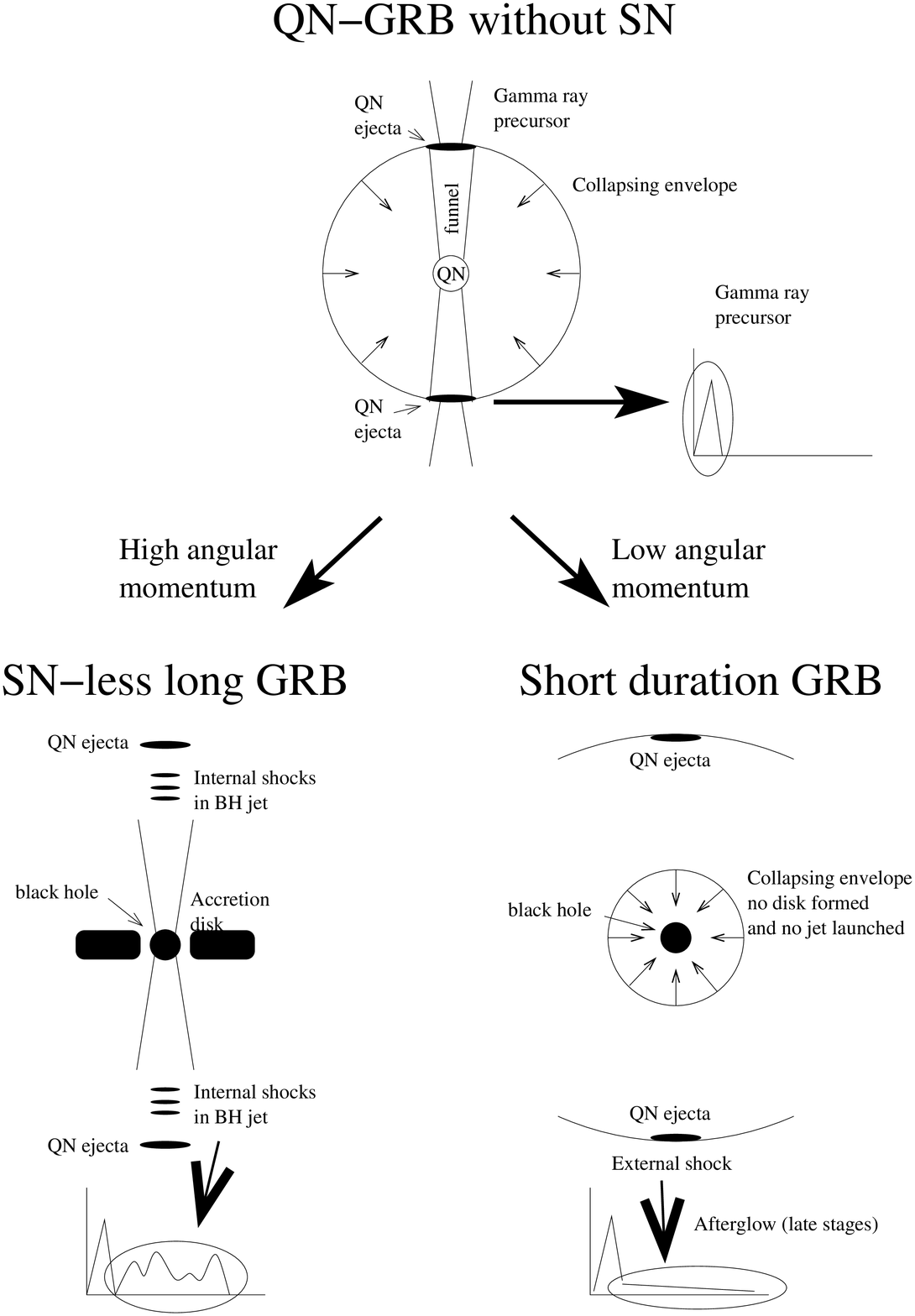}
\caption{\label{fig:snless}
Illustrated here is the case of a QN inside a failed SN (i.e. collapsing envelope).
 In this case the interaction between the denser envelope
 material and the QN ejecta would lead to a $\gamma$-ray precursor (see \S \ref{sec:thick_env}).
  The two possibles outcomes are long and short GRBs with no SN association,
   depending on the angular momentum of the progenitor.  Both cases
    result in a formation of a black hole with the higher angular
    momentum case providing an accretion disk and a jet  leading to the
    prompt GRB (see \S \ref{sec:prompt_emiss}).  The low angular momentum case consists
     on the $\gamma$-ray precursor followed only by an afterglow. }
\end{figure*}    

  \subsection{Prompt GRB emission}\label{sec:prompt_emiss}
  
As shown in Figure \ref{fig:sn}, the phase following the 
precursor phase consists of the quark star accreting
the disk material.  As shown in Ouyed et al. (2005)
whenever the quark star is heated above $T_{\rm a}\simeq$ 7.7 MeV it
will release a burst of photons with energy $\sim 3T_{\rm a}$ which can  
momentarily impede accretion, until the burst has faded at which point another
accretion episode ensues leading to another burst. 
In its simplest
 form, this episodic process (Ouyed et al. 2005)
can be responsible for creating intermittent fireballs (loaded shells with Lorentz
factor in the hundreds) eventually
 leading to  internal shocks as described by Kobayashi et al. (1997).
  Compared to  any other jet launching mechanism (e.g. from a black hole), 
the QS is able to emit far more energy for a given amount of
accreted material (Vogt et al. 2004; Ouyed et al. 2005).
 Part of the effectiveness of  our model can be attributed to the high efficiency
in which the QS converts accreted matter to radiation.

The column density of the cap is
\begin{equation}
\label{eq:column}
 N_{\rm cap} \sim \frac{m_{\rm cap}/4\pi}{56m_{\rm H}r_{\rm em}^{2}}
  \sim 2\times 10^{27}\ {\rm cm}^{-2}\ \frac{m_{\rm cap,-4}}{r_{\rm em, 12}^{2}}\ ,
\end{equation}
 with a corresponding optical depth 
\begin{equation}
\label{eq:captau}
\tau_{\rm cap} \simeq N_{\rm cap} \sigma_{\rm T} \sim 1325 \ \frac{m_{\rm cap,-4}}{r_{\rm em, 12}^{2}}\ ,
\end{equation}
 where $\sigma_{\rm T}$ is the Thompson cross-section. This 
 implies that the cap  is initially Compton optically thick to the  photons from the internal shocks occurring
underneath. 
Thus the prompt GRB phase can only be
observed as optically thin once the cap is somehow
 destroyed or pushed to a higher radius by the QS shells.

 \subsubsection{Cap acceleration and removal}  
 
In the thin envelope case, the first few shells from the QS accretion phase
  could easily remove the opaque envelope material, 
making subsequent bursts detectable by the observer. 
Alternatively, in the case of a thick envelope, the cap will be bombarded by
 many QS shells before it starts accelerating.
In general the number of collisions
with the QS shells needed to dissipate or
remove  the cap to distances large
enough to become transparent to radiation is $\sim \Gamma_{\rm f} m_{\rm cap}/\Gamma_{\rm shell} m_{\rm shell}$. That is, about   $100 $ collisions using our fiducial values.
 Equation (\ref{eq:captau}) above 
  indicates that $\tau_{\rm cap}\sim 1$ at a radius $\sim 3\times 10^{13}\ {\rm cm}\ m_{\rm cap,-4}$
   which occurs at time $\sim 1000\ {\rm s}\ m_{\rm cap,-4}$

 \subsubsection{Cap temperature and spectrum}

           An approximate equilibrium temperature for the cap can be found
from the relation, $\epsilon_{\rm r} R_{\rm QS}^{2} T_{\rm QS}^4 \sim  r_{\rm em}^{2} T_{\rm eq, cap}^{4}$,which yields
\begin{equation}
 T_{\rm eq, cap}\sim
 100~ {\rm keV}\ T_{\rm QS,10} \epsilon_{\rm r, 0.1}^{1/4} \left(\frac{R_{\rm QS,10}}{r_{\rm em,12}}\right)^{1/2}\ ,
\end{equation}
where $T_{\rm QS,10}$ and $R_{\rm QS,10}$ are the QS temperature and radius in units of 10 MeV
 and 10 km, respectively; $\epsilon_{\rm r,0.1}$ is the radiative efficiency of the internal shocks
  which we take to be $\sim 10\%$ (Kobayashi et al. 1997).
This equilibrium temperature is actually the peak temperature
because the QS heating is episodic (see Ouyed et al. 2005) and 
the temperature is only lower in between episodes.
 This quasi-continuous  supply of photons by the QS will keep
the fireball spectra close to thermal during its evolution.
 The spectra should thus consist of a blackbody in the early phase which would eventually
   evolve into an optical thin emission.
   Interestingly,  it has been suggested in the literature that a hybrid model with a thermal
  and non-thermal component can explain all type of spectral
  evolution and shapes of the observed prompt GRB emissions (e.g.
   Ryde 2005 and references therein). This is further discussed in \S \ref{sec:hybrid}.

    \subsection{Afterglow emission}
    
    The shells from the QS jet (following accretion onto the QS) colliding with the cap produce events similar
         to internal shocks between shells themselves.
    The cap provides a buffer
for the intermittent shells  
to be absorbed and subsequently form a heavy, slowly
moving ``giant'' shock (reminiscent of an external shock)
 that might be of relevance to the  afterglow activity.
 This buffer, of minimum mass $ \theta_{\rm f}^2 m_{\rm cap} 
 \sim 10^{-6}M_{\odot} \theta_{\rm f,0.1}^{2} m_{\rm cap.-4}$,  will lead
 to different type of afterglows depending on whether it is relativistic or not.

   The  slowly moving
  wall resulting from the merging of the cap and the multiple
  QS shells should absorb and emit radiation as it interacts with the surrounding
  or when it is bombarded by the energetic photons from late internal shocks (e.g.
   Staff et al. 2006b).
   Iron emission lines have been detected in the 
X-ray afterglow of few GRBs  (Piro et al. 2000; Butler et al. 2003; Reeves et al. 2002;
 Watson et al. 2002; Antonelli et al. 2000). This  might be indicative of heavy elements in the cap
  which survive nuclear disintegration due to shocks.

 \section{Discussion}
            
\subsection{SN-less GRBs in our model}
\label{sec:snlessgrbs}

If the supernova fails to explode then the consequences are twofold in our
model (see Fig.~\ref{fig:snless}). First, the QN ejecta will be subject to larger
 densities due to an infalling stellar envelope, which leads to
 higher shock efficiency $\xi_{\rm s}$ and a harder spectrum than if the
 SN had occurred.

Second, the infalling material will in most cases force the QS to turn into a black hole. 
Whereas in the SN case the outcome could be a QS
or a black hole depending on the disk and QS's initial mass, 
in the SN-less case the GRB phase is likely due to jet activity from accretion
onto a black hole.  In our QS model, as opposed to models with just a black hole, 
the black hole jet will catch up faster with the mixed ejecta/envelope since the
QN ejecta will have cleared out the more compact, dense envelope.

  \subsection{Short duration GRBs in our model}   
  \label{sec:shortgrbs}
  
   The short duration GRBs we first discuss here are necessarily
  related to star forming regions. A discussion on the second class
  (i.e. those  not associated with star forming
   regions) in our model will be presented elsewhere. 
 A short duration GRB in our model corresponds to
 the case of a low-angular momentum progenitor. In this case
 the infalling progenitor's envelope will not form a disk and will
 fall entirely onto the star, resulting in a black hole
 with no surrounding material to accrete (see Figure \ref{fig:snless}).  Simply put,
 in our model short GRBs are dominated by the precursor phase which
 will emit in the $\gamma$-ray frequency band due to the high envelope densities. 
    
We expect that the funnel's opening angle in this case will be wider 
than in cases involving disks (from angular momentum arguments).
     This implies that 
      (i) some
        short GRBs  with no SN association should be found in  star forming regions; 
         (ii)  they might be less numerous than long ones
       if low angular momentum progenitors are spars;
      (iii) they are less luminous and thus only the nearby
      one will be detectable; (iv)  their spectrum should be harder
      since the QN ejecta will interact with a more dense SN ejecta;
       (v) X-ray precursors of SN-less GRBs or the
       early phase of the prompt GRB emission in SN-less
        GRBs should resemble emission from short GRBs.

        \subsection{Dark GRBs in our model}
        \label{sec:darkgrbs}
        
Dark GRBs are defined as those that are not associated
with an optical afterglow  (e.g. Jakobsson et al. 2004)
or any afterglow emission regardless of the frequency band  (e.g. M\'esz\'aros et al. 2007).
In our model the cap as we have said provides a buffer
for the episodic shocks (from accretion onto the QS) 
to be absorbed and subsequently form an external shock
 that could in principle explain the observed afterglows.

One possible explanation using our
model is that Dark GRBs would correspond to the situations
where the interactions between the  cap and the upcoming
QS shells are reduced or nonexistent. This would be the case
if the envelope is thin in which case there is no cap or buffer, or
 if the cap is moving at relativistic speeds 
in which case the heating from the colliding shocks is diminished. 
             
           \subsection{Hypernovae as QNe signature?}\label{sec:hypernova}
           \label{sec:hypernovae}
           
            Hypernovae are energetic SNe associated with GRBs
    and are observed in the late afterfglows of
long GRBs. In our model, 
      outside the funnel where the density is above the critical density,
       the chunks will dissipate their energy entirely into heat.
        As shown in Leahy\&Ouyed
     (2008b) this results in a superluminous supernova reminiscent of
      a hypernova.  Furthermore, conditions in the expanding QN ejecta 
 are favorable for the r-process to take effect (as discussed in details in Jaikumar et al. 2007).
Hence, additional heavy elements will be deposited in the expanding envelope
 as the QN ejecta reaches and mixes with the envelope.

\subsection{Optical flashes and X-ray precursors}
\label{sec:opticalemission}

 Traditionally, it has been suggested that the optical emission  could be produced by the reverse shock 
  emission or could arise from the internal shock emission (e.g. Sari \& Piran 1999). 
 Observationally, there appears to be two cases of  prompt optical emission: 
 \begin{itemize}
 \item  For GRB990123 and recently discovered GRB060111b, the 
optical flashes were uncorrelated with the prompt gamma-ray emission, which suggests that 
the optical emission and gamma-ray emission should have different origin (e.g. Klotz et al. 2006). 

\item For GRB041219a, its optical flash was correlated 
with the gamma-ray emission (Vestrand et al. 2005; Blake et al. 2005), and for GRB050904, 
a very bright optical flare was temporally coincident with an X-ray flare (B\"oer et al. 2006; see
also Wei 2007), 
which implies that for these two GRBs there should be a common origin for the optical and 
high energy emission.  
\end{itemize}
 More recently, the near-simultaneous optical and $\gamma$-ray  emission in GRB 080319B
 has re-opened this debate. Kumar\&Panaitescu (2008) argue for a Synchrotron self Compton (SSC) origin
  while recent work (Zou, Piran, \& Sari 2008) argue for physically separated
   emission regions (e.g. $\gamma$-rays  from internal shocks and optical flash  from external shock
emission).  
   
    In our model, the optical emission originates when the chunks
     are heated during their passage through the thin envelope (see \S 3.1.1), and
      X-ray emission when the chunks dissipate further out at the density
       inversion (high metallicity stars;  Petrovic et al. 2006).
  The optical and X-ray emission are produced at different radii with durations dominated by
    geometry as given in Eq.(6) but both composed of short pulses from the $\sim 10^3$-$10^7$ chunks.  
  The prompt GRB emission 
   is from internal shocks in the outflow launched during accretion onto the QS (see \S 3.2), 
     physically separated from the optical emission.  The delay of the optical from the QN event
     can be shorter or longer than the corresponding GRB delay  depending on the Lorentz factors
       and emission distances.

\subsection{X-ray flares}
\label{sec:xflares}

X-ray flares are frequently observed in the early X-ray afterglow of GRBs
(e.g. Burrows et al. 2005; Chincarini et al. 2007). In Staff, Ouyed,
\& Bagchi  (2007) a possible explanation for these X-ray flares with an accreting quark
star as the GRB inner engine is given. A quark star can accrete a
maximum of about $0.1 M_{\odot}$ before collapsing to a black hole. Hence the
quark star as GRB inner engine can last for about a thousand seconds. If there is
still matter left to accrete once the black hole is formed, a new jet is
formed from the accretion onto the black hole (DeVilliers, Staff, \& Ouyed
 2005). Staff et al. (2007) proposed
that the interaction between the jet from the black hole and the jet from
the quark star could make internal shocks and thereby produce X-ray flares.
Furthermore, when the black hole jet (or more massive parts of the quark
star jet) interacts with the external shock, the shock will be reenergized
and a ``bump'' can result. Internal shocks within the black hole jet itself
can also occur, giving rise to more flares; see Figure 5 in Staff et al. (2007)
 for the resulting lightcurves.

The accretion rate onto a black hole is likely very high ($\dot{M}_{\rm
BH}\sim0.1-1 M_{\odot} \ {\rm s}^{-1}$), meaning the black hole phase will be rather
short. The flares created by interactions between the black hole jet and the
quark star jet therefore have to occur within about a thousand seconds,
whereas the activity caused by interaction between the jets and the external
shock can occur later. Flares can occur earlier if the quark star collapses 
to a black hole sooner than after a thousand seconds.

\subsection{The two components}
\label{sec:hybrid}

  From equation (\ref{eq:column})  we estimate the expanding 
    cap to becomes optically thin to photons from the internal shocks, at a radius of about
     $r_{\rm therm.}  \sim 3\times 10^{13}\ {\rm cm}\ m_{\rm cap,-4}$
     at which point the temperature of the thermal component becomes
     $T_{\rm eq}\sim 10\ {\rm keV}\ T_{\rm QS,10} R_{\rm QS,10}^{1/2}/m_{\rm cap,-4}^{1/4}$.
     The emission would thus remain   thermal for a duration of $t_{\rm therm.}\sim 
     r_{\rm therm.}/(2\Gamma_{\rm i}^2 c)\sim 10\ {\rm s}\ m_{\rm cap,-4}^{1/2}/\Gamma_{\rm i,10}^2$. The synchrotron radiation from the
     subsequent internal shocks  will then dominate the spectrum in the later stages, $r>r_{\rm therm.}$,  of the
     prompt GRBs;  the QS will continue to accrete until 
     the accretion disk is consumed or the QS turns into a black hole (Ouyed et al. 2005;
     Staff et al. 2006b). Whether the two-components presumably inherent
      to some GRBs (Ryde 2005) is an indication of 
    the envelope-shells interaction as described in our model remains to be confirmed.
     For completeness, however, we should note that accreting quark stars
     could in principle result, as we have said above,  
     from QNe going off in isolation in which case  the standard
       internal shocks scenario, involving no intervening  envelope, applies (Ouyed et al. 2005).

\section{Conclusion}
\label{sec:conclusion}

 In this paper, we explore the case of a QN going off inside a collapsar.
  We find that  the interaction
  between the iron-rich chunks  from the QN ejecta and the collapsar envelope leads to features indicative of those observed in Gamma Ray Bursts.  
These features include: (i) precursor  activity (optical, X-ray, $\gamma$-ray), 
 (ii) prompt  $\gamma$-ray emission, and (iii) afterglow emission.
Although the presented model is based on physical arguments, 
most of these are in reality more complicated and so would require more detailed studies.
For example,
the launching of the outer layers of the neutron star during the QN 
  is a challenging  process to study, and involves energy transfer, core-bounce, 
   generation of a shock wave, including cooling processes, and subsequent ejection. 
    We have assumed simple conditions for the ejecta immediately after
     the QN such as a single Lorentz factor.  A range of Lorentz
    factors  would still result in the outermost  shell of the ejected material 
     interacting with the progenitor envelope as
     we have described here. Shells with lower Lorentz factors would interact
      later in a similar manner and
      would lead to more complex interaction with the envelope.
      Another important aspect of our model that requires
      further studies is the process of clumping, crystallization, and breakup
       of the ejecta, which would require better knowledge of the
       ambient conditions surrounding the ejecta.
      Despite our simplifying assumptions, we feel that our model captures the basic envelope
      interaction physics  and 
       provides interesting features  with possible applications to GRBs. 

\begin{acknowledgements}
 This work is supported by an operating grant from the Natural Research Council of Canada (NSERC).
\end{acknowledgements}

\end{document}